\documentclass[onecolumn,floatfix,superscriptaddress,a4paper,
               showpacs,showkeys,nofootinbib,preprint]{revtex4}

\usepackage{epsfig}
\usepackage{amssymb,latexsym,amsmath}
\newcommand{\eq}[1]{\begin{align} #1 \end{align}}

\begin{document}

\title{Crossover to Cluster Plasma in the Gas of Quark-Gluon Bags}

 \author{Viktor V. Begun}
 \affiliation{Bogolyubov Institute for Theoretical Physics,
 Kiev, Ukraine}
 \affiliation{Frankfurt Institute for Advanced Studies, Frankfurt,
 Germany}

 \author{Mark I. Gorenstein}
 \affiliation{Bogolyubov Institute for Theoretical Physics,
 Kiev, Ukraine}
 \affiliation{Frankfurt Institute for Advanced Studies, Frankfurt, Germany}

 \author{Walter Greiner}
 \affiliation{Frankfurt Institute for Advanced Studies,
 Frankfurt, Germany}

\begin{abstract}
 We study  a smooth crossover transition in the gas
 of quark-gluon bags. The equation of state at high temperature
 is that of the quark-gluon plasma. However, the system consists of
 the bags with finite volumes which are defined by the model parameters
 of the mass-volume bag spectrum.  Possible structures in
 this {\it cluster} quark-gluon plasma are classified.
\end{abstract}



\maketitle

\section{Introduction}
A connection of the bag model \cite{Bag} with statistical
description of strongly interacting matter at high energy density
has a long history \cite{Hagedorn}. The possibility of phase
transitions in the gas of quark-gluon bags was demonstrated for
the first time in Ref.~\cite{pt}. Further studies allowed to
obtain the 1st, 2nd, and higher order transitions
\cite{pt1,GGG,Zakout,Bugaev,Bessa}. A possibility of no phase
transitions was also pointed out \cite{pt1,GGG}. Recently it was
suggested \cite{Ferroni} to model a smooth crossover transition by
the gas of quark-gluon bags. Inspired by this suggestion we study
in more details the high temperature behavior of the system of
quark-gluon bags in case of the crossover. Note that the present
and future experimental facilities such as RHIC and LHC produce
strongly interacting matter in the crossover region of the QCD
phase diagram \cite{Aoki}. Moreover, the location of the
(tri)critical point that ends the line of the 1st order phase
transition is unknown \cite{Stephanov}. It can happen that
nucleus-nucleus collisions at SPS and FAIR also partially enter
the crossover region.

Section~\ref{sec-QG-bags} gives a short overview of the gas of
bags with excluded volume. In Section~\ref{sec-g-d} the phase
diagram of the model parameters in the mass-volume bag spectrum is
considered. The average volume of the bags is calculated in the
high temperature limit. In Section~\ref{sec-Summary} a short
summary of the results is presented.

\section{Gas of Quark-Gluon Bags}\label{sec-QG-bags}
The partition function for gas of
quark-gluon bags is the following \cite{pt}:
\begin{align}\label{Z}
 Z(V,T)~ &=  ~\sum_{N=0}^{\infty}
       \frac{1}
       {N!}~
       \prod_{i=1}^N
       \int dm_i dv_i\;\rho(m_i,v_i)\;\phi(T,m_i)
       \left(V-\sum_{j=1}^N v_j\right)^N
       \theta\left(V-\sum_{j=1}^N v_j\right)~,
\end{align}
where $V$ and $T$ are the system volume and temperature
respectively, $N$ is number of bags,
$m_i$, $v_i$ are  mass and  proper volume of  $i$-th bag,
$\rho(m_i,v_i)$ is the mass-volume spectrum of bags that will be
specified later. The $\phi(T,m_i)$ in Eq.~(\ref{Z}) equals:
\begin{align}\label{phim}
\phi(T,m_i)~  \equiv ~\frac{1}{2\pi^2}~\int_0^{\infty}k^2dk~
\exp\left[-~\frac{(k^2~+~m_i^2)^{1/2}}{T} \right]
~=~\frac{m_i^2T}{2\pi^2}~K_2\left(\frac{m_i}{T}\right)~,
\end{align}
and it has the physical meaning of the particle number density in
Boltzmann ideal gas, $K_2$ in Eq.~(\ref{phim}) is the modified
Bessel function.
The baryonic number and other conserved charges are assumed to be
equal to zero. The equation of state can be most easily studied
with the help of the Laplace transform:
\begin{align}\label{Zs}
 \hat{Z}(T,s)~&\equiv\int_0^{\infty}dV\exp(-sV)~Z(V,T)
~=~\left[~s~-~f(T,s)\right]^{-1}~,
\end{align}
where
\begin{align}\label{fs}
  f(T,s)~&=~  \int_0^{\infty} dmdv\, \exp(-vs)
~\rho(m,v)\,\phi(T,m) ~.
\end{align}

In the thermodynamic limit, $V\rightarrow\infty$, the partition
function behaves as $Z(V,T)\cong \exp\left[p\,V/T \right]$, where
$p(T)$ is the system pressure. An exponential increase of $Z(V,T)$
on $V$ generates the singularity $s^*$ of the function
$\hat{Z}(T,s)$ in variable $s$.
Consequently, one can calculate the pressure
knowing only the position of the farthest-right singularity $s^*$:
%
%
 \begin{align}\label{p-s}
 p(T)~=~T~\lim_{V\rightarrow\infty}\frac{\ln
 Z(V,T)}{V}~=~T~s^*(T)~.
 \end{align}
There is a pole singularity $s^*=s_H$ of $\hat{Z}(T,s)$
calculated from the transcendental equation:
%
 \begin{align}\label{sH}
 s_H(T) ~=~f(T,s_H(T))~.
 \end{align}
Another singular point of  $\hat{Z}(T,s)$
denoted as
$s_Q(T)$ emerges due to a singularity of the function $f(T,s)$
itself. The system pressure takes then the form
\cite{pt}:
 \begin{align}\label{ps*}
 p(T)~=~Ts^*(T)~=~T\cdot max\{s_H(T),s_Q(T)\} ~,
 \end{align}
i.e. the farthest-right singularity $s^*(T)$ of $\hat{Z}(T,s)$
(\ref{Zs}) can be either the pole singularity $s_H(T)$ (\ref{sH})
or the $s_Q(T)$ singularity of the function $f(T,s)$ (\ref{fs})
itself. The mathematical mechanism for possible phase transition
(PT) in the gas of quark-gluon bags is the ``collision'' of the
two singularities, i.e. $s_H(T)=s_Q(T)$ at the PT temperature
$T=T_C$.

The crucial ingredient of the  model  is the form of the
mass-volume spectrum  $\rho(m,v)$.
In the case of a bag filled with the non-interacting massless
quarks and gluons\footnote{This picture is reasonable in the
region where both the mass of the bag, $m$, and the volume of the
bag, $v$, are large. A general form of the mass-volume spectrum
function should contain also the low-lying hadron states.
However, the type of the PT or the high temperature behavior when
the PT is absent are not sensitive to presence of these
hadron-like states.
} one finds \cite{pt,pt1,GGG}:
\begin{align}\label{rhomv}
 \rho(m,v)~\simeq
 ~C~v^{\gamma}(m-Bv)^{\delta} ~\exp\left[\frac{4}{3}~\sigma_Q^{1/4}~
 v^{1/4}~(m-Bv)^{3/4}\right]~,
\end{align}
where $C$, $\gamma$, $\delta$ and $B$, the so-called bag
constants, are the model parameters, and $\sigma_Q=95\pi^2/60$
is the Stefan-Boltzmann constant counting gluons (spin, color) and
(anti-)quarks (spin, color and $u$, $d$, $s$-flavor) degrees of
freedom inside the bag. This is the asymptotic expression assumed
to be valid for a sufficiently large volume and mass of a  bag,
$v>V_0$ and $m>Bv+M_0$. The validity limits can be estimated to be
$V_0\approx
1$~fm$^3$ and $M_0\approx 2$~GeV \cite{GGG}.

The integral over mass  in Eq.~(\ref{fs}) can be calculated by the
steepest descent estimate and one finds:
\begin{align}\label{fQs}
 f(T,s)
 &~\simeq~ C\int_{V_0}^{\infty}dv~v^{\gamma} \exp{\left[\,-v\left(s-s_Q\right)\right]}
   \int_{M_0}^{\infty}dm\, (m-Bv)^{\delta}
   \left(\frac{mT}{2\pi}\right)^{3/2}
   \exp\left[\,-\,\frac{(m-\overline{m})^2}{8\sigma_Q vT^5}\right]
   \nonumber \\
&~\simeq~
 u(T)\int_{V_0}^{\infty}dv~v^{2+\gamma
+\delta}~\exp\left[-v\left(s~-s_Q(T)\right)\right]~,
 \end{align}
where $\overline{m}=v(\sigma_QT^4+B)$,~ $u(T)=C\pi^{-1}
\sigma_Q^{\delta+1/2}~T^{4+4\delta}
~\left(\sigma_QT^4+B\right)^{3/2}$ and
 \begin{align}\label{sQ}
 s_Q(T)~\equiv~ \frac{1}{3}~\sigma_Q~T^3~-~\frac{B}{T}~.
 \end{align}


%
%
\section{The $\gamma-\delta$ phase diagram}\label{sec-g-d}
%
%

\begin{figure}[h!]
 \epsfig{file=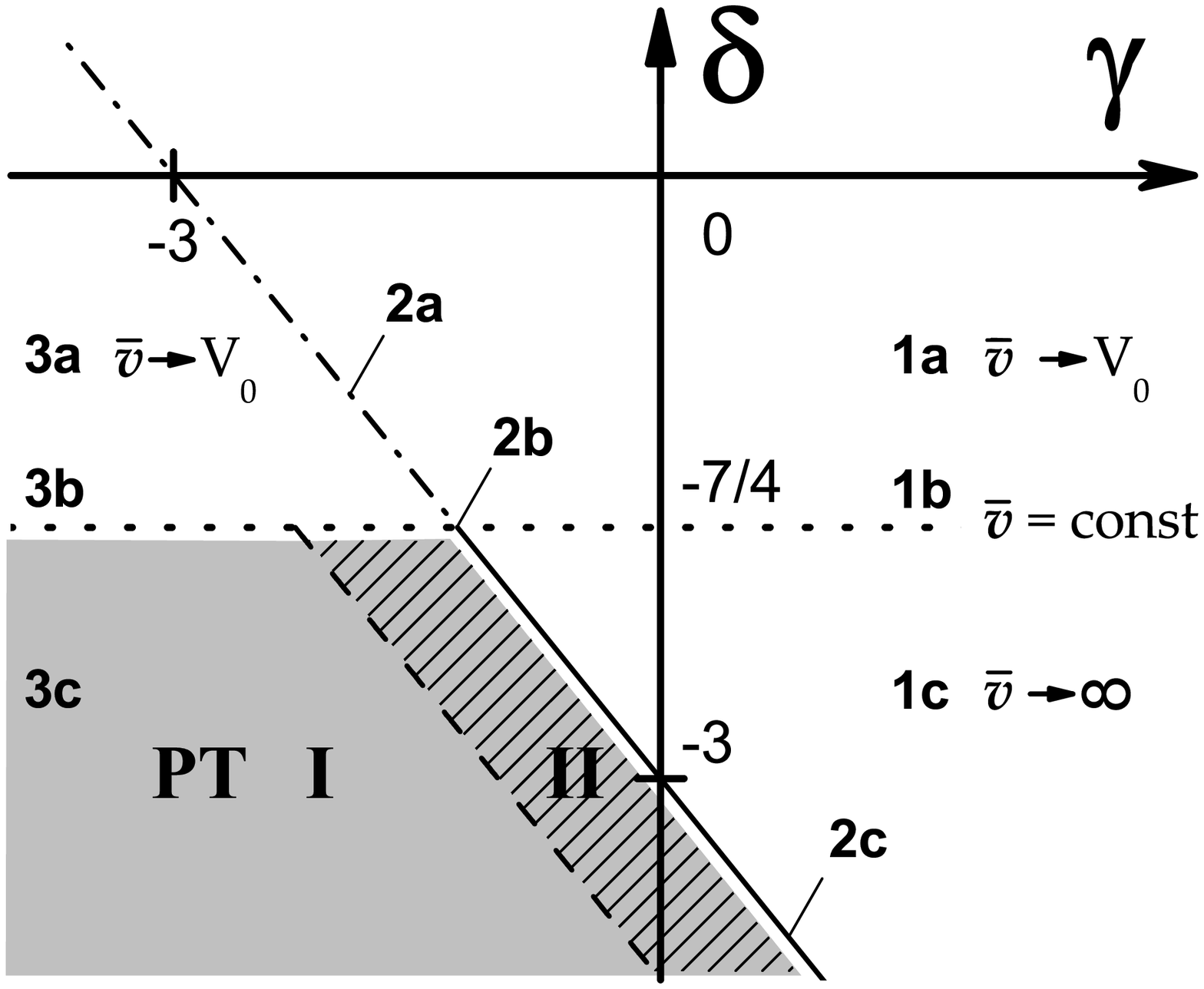,width=0.5\textwidth}
 \epsfig{file=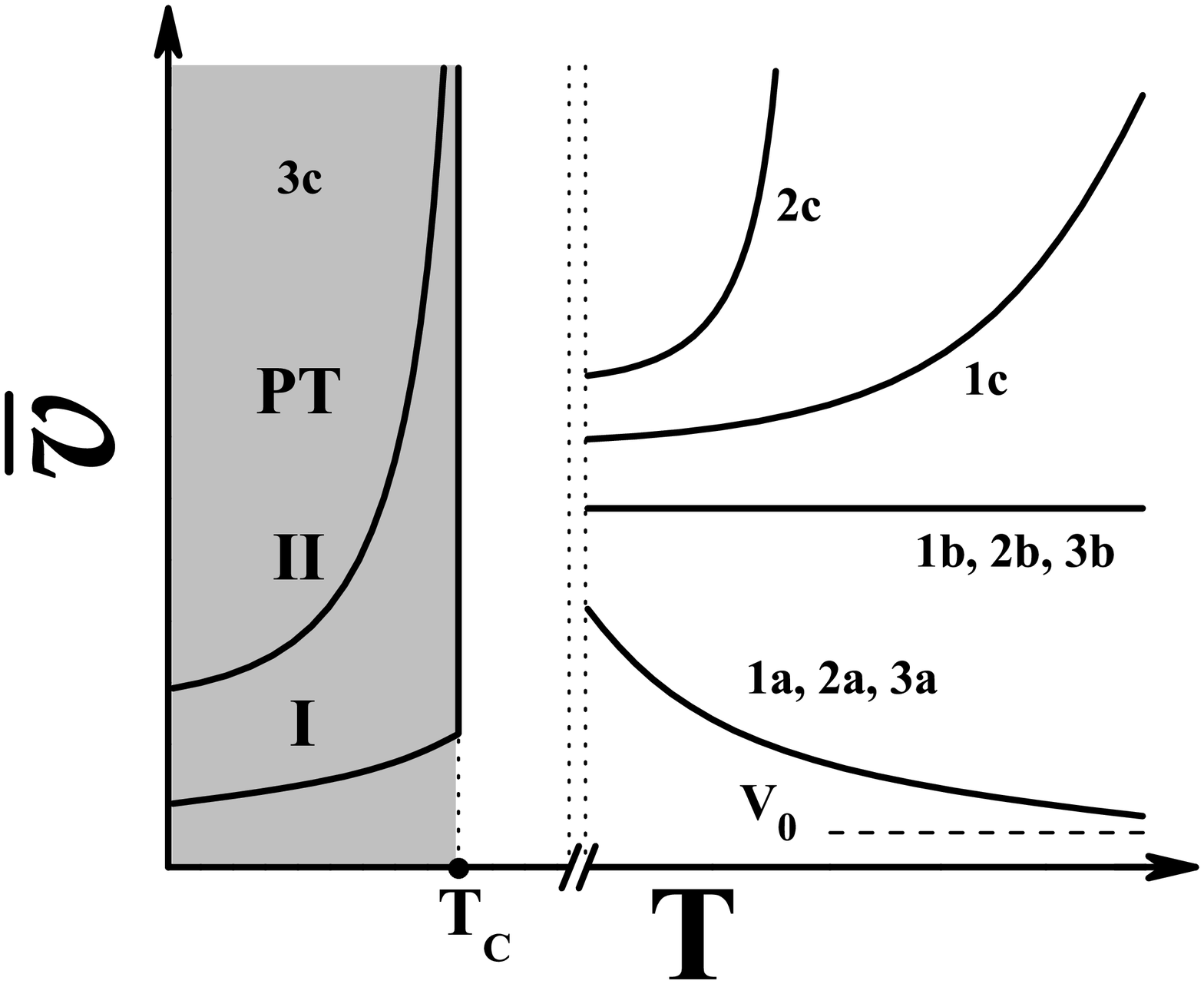,width=0.49\textwidth}
 \caption{{\it Left:} The $\gamma-\delta$ phase diagram
 for the gas of quark-gluon bags. {\it Right:}
 The schematic view of temperature dependence
 of the average volume of the bag in different regions of $\gamma-\delta$ phase
 diagram. See text for details.\label{fig-gd}}
\end{figure}

The model parameters $\gamma$ and $\delta$ define the presence,
location and order of the PT in the gas of quark-gluon bags. This
is illustrated by the $\gamma-\delta$ phase diagram in
Fig.~\ref{fig-gd} ({\it left}).

Most interest in the previous studies
\cite{pt,pt1,GGG,Zakout,Bugaev,Bessa} was devoted to PTs.
Different order PTs are discussed in details in Ref.~\cite{GGG}.
We concentrate here on the situations when PTs are absent.
%
%
%
%
Eq.~(\ref{sH}) for $s_H(T)$ can be written as follows:
 \eq{\label{s-high-T}
 s_H ~=~ u(T)
  ~ \int_{V_0}^{\infty}dv~v^{a-1}~
    \exp\left(-v~\Delta s
    \right)~\propto~ T^{10+4\delta}~\left( \Delta s\right)^{-a}~\Gamma(a,V_0 \Delta s)~,
 }
where $a\equiv \gamma +\delta +3$,~ $\Delta s(T)\equiv s_H-s_Q$,
and $\Gamma(a,b)$ is the incomplete Gamma-function. The function
$f(T,s)$
has the singular point $s=s_Q$, and
$f(T,s_Q)$ can be either finite or infinite depending on the value
of $\gamma+\delta$:
\eq{\label{a}  1)~~
 \gamma+\delta > -3~, ~~~~
%
 2) ~~ \gamma+\delta= -3~, ~~~~ 3) ~~
  \gamma+\delta < -3~.
 }

\begin{figure}[h!]
 \epsfig{file=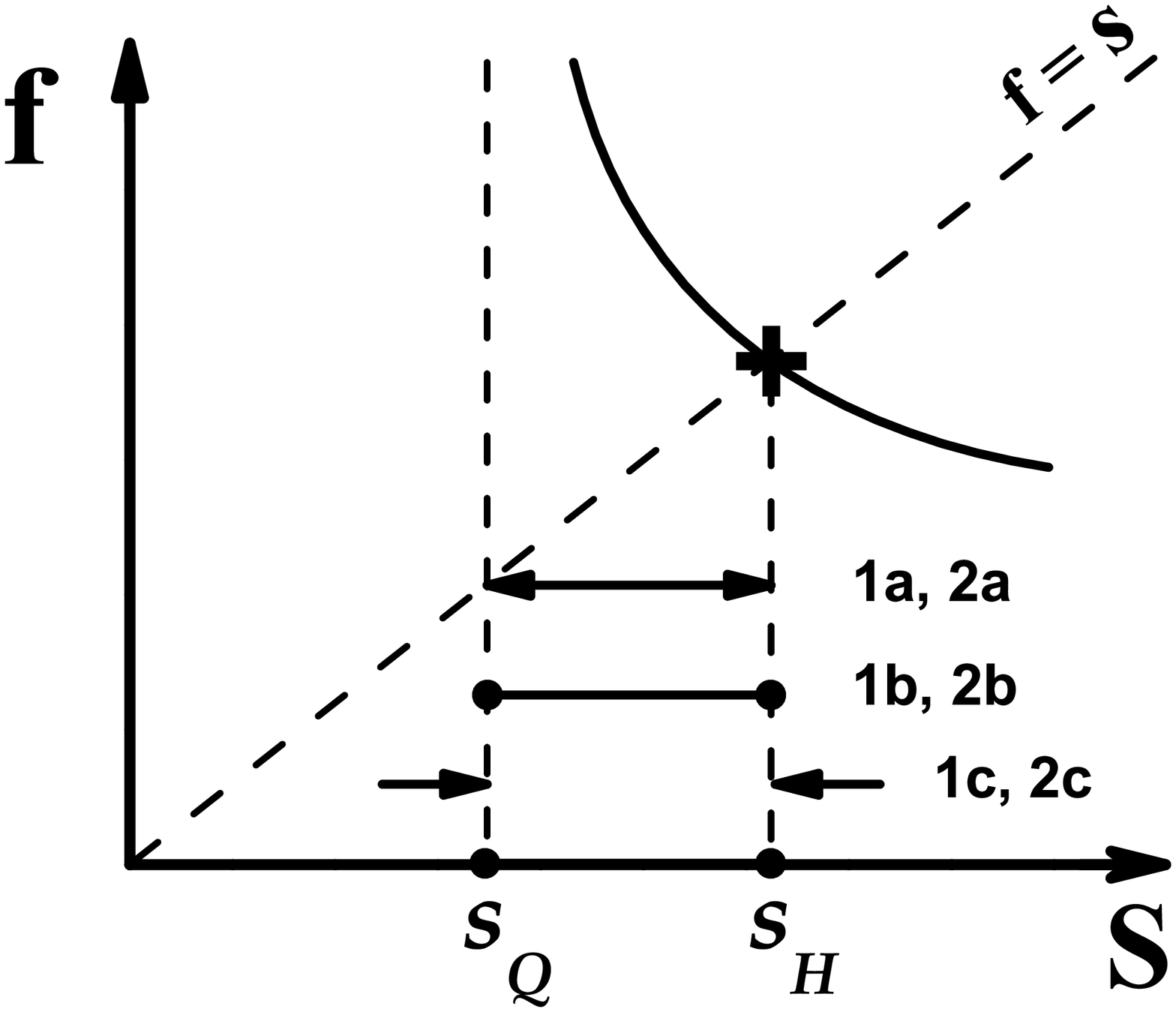,width=0.32\textwidth}\hspace{0.2cm}
 \epsfig{file=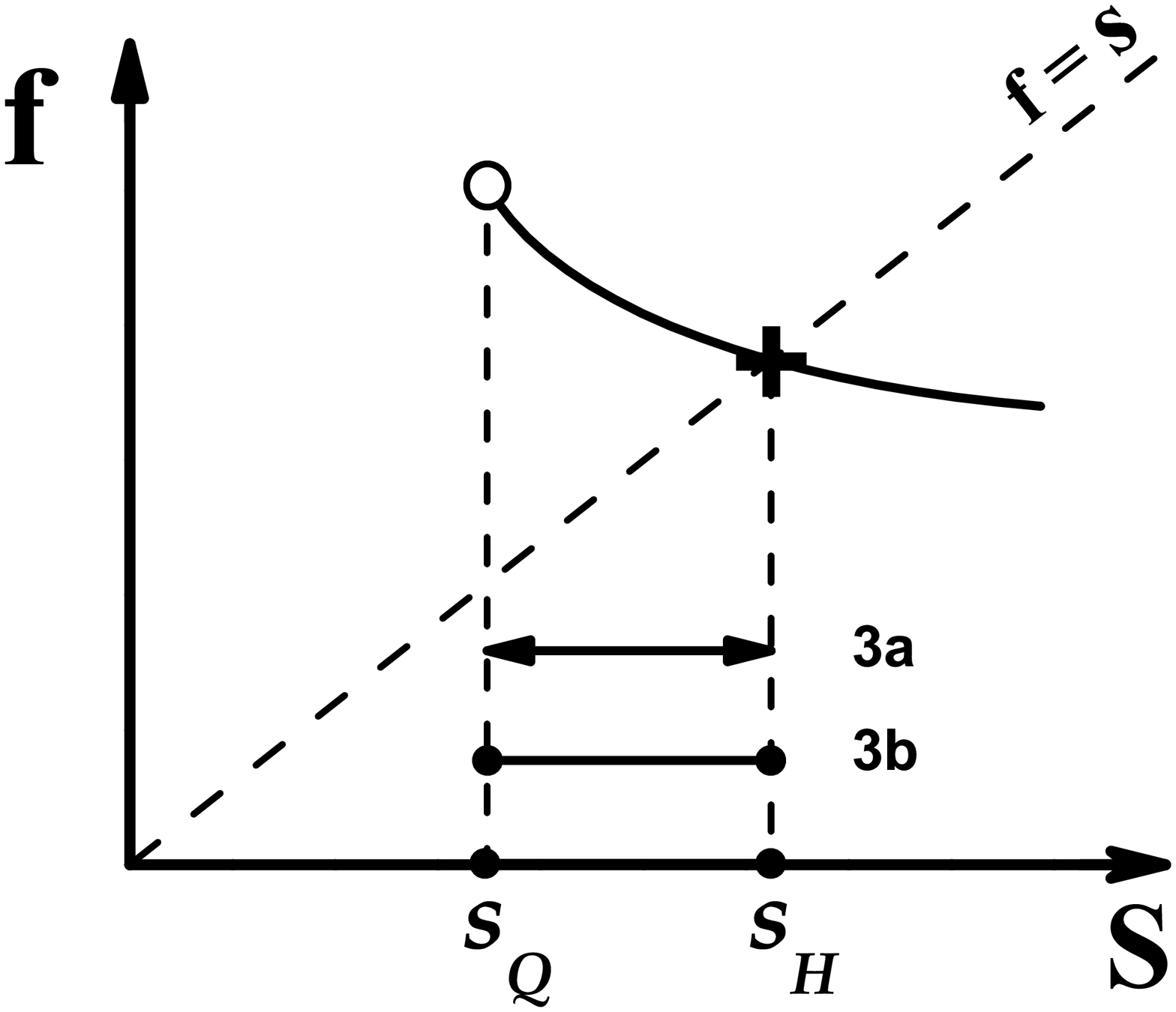,width=0.32\textwidth}\hspace{0.2cm}
 \epsfig{file=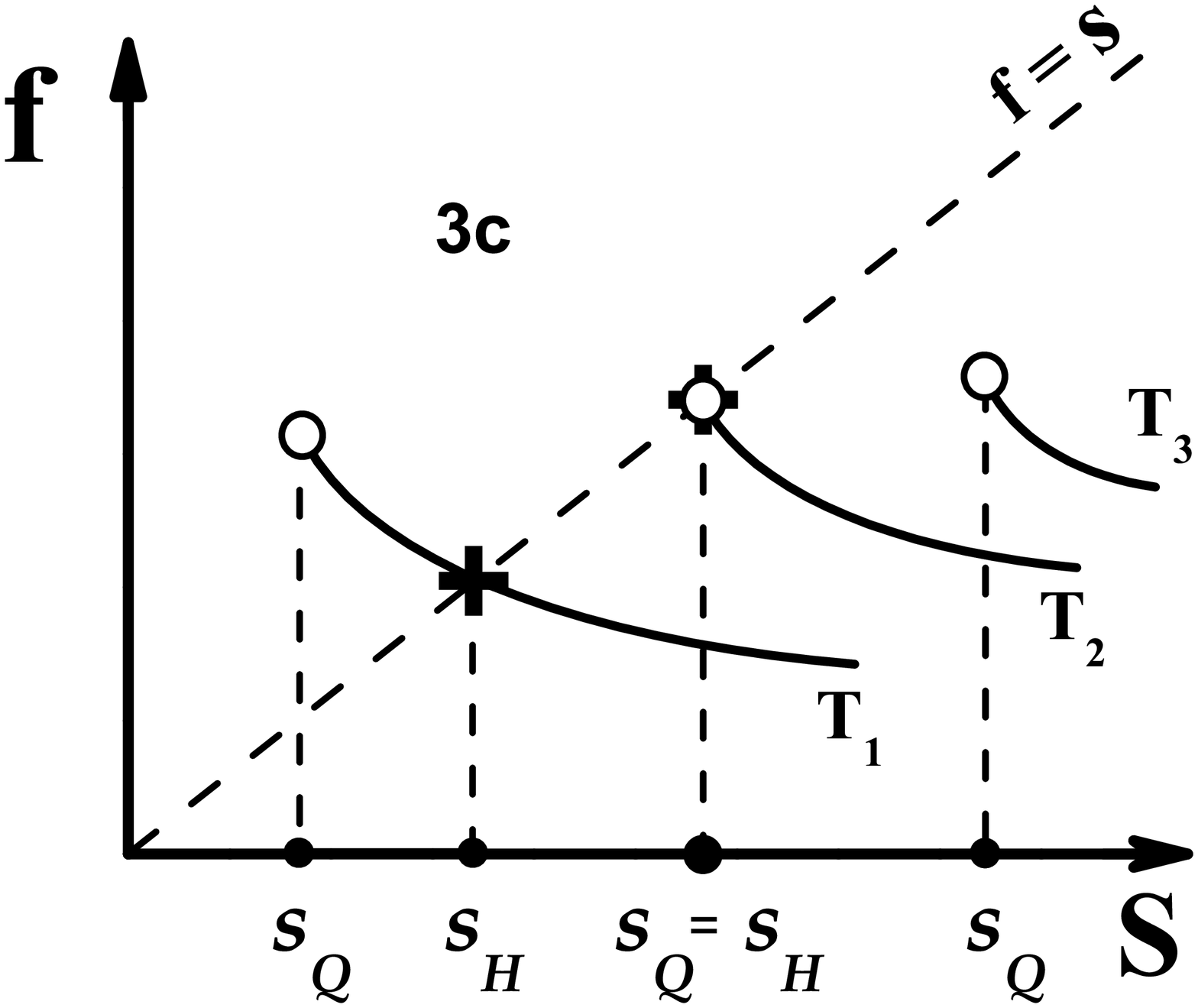,width=0.32\textwidth}
 \caption{ The solid lines present the dependence of $f$
on $s$ at fixed temperatures. The pole singularity $s_H$ and
singularity $s_Q$ are denoted by crosses and circles,
respectively. {\it Left:} The cases 1 and 2 in Eq.~(\ref{a})
correspond to $f(T,s_Q)=\infty$. {\it Middle:} The cases 3a and 3b
in Eqs.~(\ref{a},\ref{d}) are shown. They correspond to
$f(T,s_Q)>s_Q$ at all $T$. {\it Right:}
The case 3c with $T_1<T_2=T_C<T_3$. It leads to the ``collision''
of two singularities $s_H=s_Q$ at the PT temperature $T_C$.
\label{fig-fs1}}
\end{figure}

The phase transition never exists for $\gamma+\delta\geq -3$
because of $f(T,s_Q)=\infty$, and in this case the solution
$s_H(T)$ of Eq.~(\ref{s-high-T}) always corresponds to $s_H>s_Q$,
see Fig.~\ref{fig-fs1} {\it left}. For $\gamma+\delta < -3$ the
existence of a PT depends on the value of parameter $\delta$.
There are three distinct cases:
\eq{\label{d} a)~~\delta>-7/4~,~~~~b)~~ \delta=-7/4~,~~~~ c)~~
\delta < -7/4~.
}

The PTs take place in the system of quark-gluon bags in the only
case 3c, i.e. if $\gamma+\delta<-3$ and $\delta<-7/4$.  This is
the lower left corner of the $\gamma-\delta$ phase diagram marked
as 3c and shown by grey color in Fig.~\ref{fig-gd} {\it left}. The
region with 1st and 2nd or higher order PTs are marked as I and II
correspondingly.

 The average volume of the quark-gluon bag can be
calculated as:
 \eq{\label{v(T)}
 \overline{v}(T)~=~
\frac{\int  dvdm~v~\rho(m,v)~\phi(T,m)~\exp(-s^*v)}{\int dvdm~
\rho(m,v)~\phi(T,m)~\exp(-s^*v)}
~ \cong ~ \frac{1}{\Delta s(T)}\;\frac{\Gamma(a+1,V_0\,\Delta
s(T))}{\Gamma(a,V_0\,\Delta
 s(T))}~.
 }
It can be proven that $s_H \sim T^3$ for $T\rightarrow\infty$.
Using the asymptotic expansion for incomplete $\Gamma$-function
(see Appendix) one obtains from Eqs.~(\ref{s-high-T}),
(\ref{v(T)}) the behavior of average volume $\overline{v}(T)$ of
the quark-gluon bag at large temperature $T$:
\eq{\label{1a}
  1a,~2a,~3a~: ~~~~
&  \Delta s(T) \;\sim\;\,\ln T~
 \rightarrow\,\infty\;,~~~~ && \overline{v}(T)\rightarrow V_0~,\\
%
%
\label{1b}
  1b,~2b,~3b~:~~~~
&  \Delta s(T) \;\cong\;const~>~ 0\;~~~~ && \overline{v}(T)\rightarrow const~,\\
%
%
%
%
%
%
%
1c~: ~~~~
& \Delta s(T) \;\sim\;T^{(7+4\delta)/a}\;\rightarrow\; 0\;,~~~~ &&
\overline{v}(T)\sim T^{-(7+4\delta)/a}\;\rightarrow\; \infty \;,
 \label{1c}
 \\
 2c~: ~~~~
&\Delta s(T) \;\sim\;\exp\left(-~T^{-7-4\delta}\right)
     \;\rightarrow\; 0\;,~~~&&\overline{v}(T)
\;\sim\;\exp\left(T^{-7-4\delta}\right)
     \;\rightarrow\; \infty \;.\label{2c}
}
The results (\ref{1a}-\ref{2c}) for $\overline{v}(T)$ are
schematically presented in Fig.~\ref{fig-gd} {\it right}. Note
that if the phase transition takes place at $T=T_C$ (case 3c) the
$\overline{v}(T)$ becomes infinite at $T>T_C$ in the thermodynamic
limit. This is also shown in Fig.~\ref{fig-gd} {\it right}.

%
\section{Summary}\label{sec-Summary}
%
%
In this paper we have studied the gas of quark-gluon bags at high
temperature $T$. The mass-volume spectrum function $\rho(m,v)$ for
the quark-gluon bags is taken in the form (\ref{rhomv}). The
behavior of the system
depends crucially on the values
of the $\gamma$ and $\delta$ parameters in Eq.~(\ref{rhomv}).
The special regions of the $\gamma-\delta$ phase diagram defined
by Eqs.~(\ref{a},\ref{d}) lead to different behavior of the gas of
the quark-gluon bags.  The pressure $p$ and energy density
$\varepsilon$ for different values of $\gamma$ and $\delta$ have
the same asymptotic behavior at high temperature, $p\cong
\sigma_QT^4/3$ and $\varepsilon\cong \sigma_Q T^4$. This
corresponds to the equation of state of non-interacting quarks and
gluons inside the bags, i.e. {\it ideal} quark-gluon plasma (QGP).
However, the average volume of the bag $\overline{v}(T)$
(\ref{1a}-\ref{2c}) and average mass $\overline{m}(T)\cong
\overline{v}(\sigma_QT^4+B)$ have rather different behavior in
different regions of the $\gamma-\delta$ phase diagram. If the
system of quark-gluon bags has no phase transition the
$\overline{v}(T)$ remains finite at high temperature. Such a {\it
cluster} QGP can be rather different from the {\it ideal} QGP
despite of the similar to that equation of state. The kinetic
properties of the {\it cluster} QGP, e.g. the shear and bulk
viscosity, may deviate strongly from those in the quark-gluon gas.

\vspace{0.5cm} {\bf Acknowledgments} 

This work was in part supported by the Program of Fundamental
Researches of the Department of Physics and Astronomy of National
Academy of Sciences, Ukraine. V.V. Begun thanks the Alexander von
Humboldt Foundation for the support.

%
%
\appendix
\section{}\label{app-A}
%
The incomplete gamma function $\Gamma(a,x)$ (see, e.g.,
Ref.~\cite{Abr}) has the asymptotic expansions at $x\rightarrow
\infty$,
 \eq{\label{G-inf}
 \Gamma(a,x\rightarrow\infty)
 \;=\; x^{a-1}e^{-x} \left(1 \;+\; \frac{a-1}{x} \;+\; O(x^{-2})
 \right)\;,
 }
%
and $x\rightarrow 0$,
 \eq{\label{G-0}
 \Gamma(a,x\rightarrow 0)
 & \;=\; \Gamma(a)\;-\;\frac{x^a}{a}\left(1\;+\;O(x)\right)\;,
        &&a\neq -n\;,
 \\\nonumber
 & \;=\; \frac{(-1)^n}{n!}\left(\psi(n+1)\;-\;\ln x\right)
   \;+\; \frac{x^{-n}}{n}\left(1\;+\;O(x)\right)\;, &&a=
   -n\;,
 }
where $\psi(z)$ is the logarithmic derivative of the gamma
function,  $\psi(z)=\Gamma'(z)/\Gamma(z)$. Eq.~(\ref{G-0}) gives
for $a=0,\,-1,\,-2$:
 \eq{
 \Gamma(a,x\rightarrow 0)
 & \;=\; -\ln x \;+\; O(1)\;, &&a=0
 \\
 & \;=\; \ln x \;+\; \frac{1+O(x)}{x}\;, &&a=-1\;,
 \nonumber\\\nonumber
 & \;=\; \frac{1+O(x)}{2x^2}\;, &&a=-2\;.
 }
%

%
%

\end{document}